\PassOptionsToPackage{table,xcdraw}{xcolor}

\documentclass[sigconf,nonacm,natbib=false,screen=true]{acmart}
\usepackage[table]{xcolor}
\usepackage{pgf}
\pdfoutput=1 

\usepackage{subcaption}
\usepackage{enumitem}
\usepackage{makecell,multirow}
\usepackage{titlesec}
\usepackage{tabularx}
\usepackage{color, colortbl}

\newcommand\ColRowGB[1]{\pgfmathsetmacro\compH{0.5}%
  \pgfmathsetmacro\compS{#1 * 0.2}\pgfmathsetmacro\compB{1}%
  \edef\x{\noexpand\cellcolor[hsb]{\compH,\compS,\compB}{#1}}\x}

\newcommand\ColRowAmp[1]{\pgfmathsetmacro\compH{1}%
  \pgfmathsetmacro\compS{(#1 -0.9) * 0.16}\pgfmathsetmacro\compB{1}%
  \edef\x{\noexpand\cellcolor[hsb]{\compH,\compS,\compB}{#1}}\x}

\newcommand\ColSeCo[1]{\pgfmathsetmacro\compH{0.6}%
  \pgfmathsetmacro\compS{(#1 -0.1) * 0.1}\pgfmathsetmacro\compB{1}%
  \edef\x{\noexpand\cellcolor[hsb]{\compH,\compS,\compB}{#1}}\x}

\newcommand\ColRaCo[1]{\pgfmathsetmacro\compH{0.6}%
  \pgfmathsetmacro\compS{(#1 -0.1) * 0.1}\pgfmathsetmacro\compB{1}%
  \edef\x{\noexpand\cellcolor[hsb]{\compH,\compS,\compB}{#1}}\x}

\newcommand\ColSeWa[1]{\pgfmathsetmacro\compH{0.3}%
  \pgfmathsetmacro\compS{(#1 -2.0) * 0.04}\pgfmathsetmacro\compB{1}%
  \edef\x{\noexpand\cellcolor[hsb]{\compH,\compS,\compB}{#1}}\x}

\newcommand\ColRaWa[1]{\pgfmathsetmacro\compH{0.3}%
  \pgfmathsetmacro\compS{(#1 -2.0) * 0.04}\pgfmathsetmacro\compB{1}%
  \edef\x{\noexpand\cellcolor[hsb]{\compH,\compS,\compB}{#1}}\x}

\newcommand{\myepsilon}{ \cellcolor[hsb]{0.5,0.01,0.98}$\varepsilon$ }

\usepackage[backend=biber,backref=false,doi=false,eprint=false,isbn=false,url=false,
  maxbibnames=99,style=numeric,sortcites]{biblatex}
\AtEveryBibitem{
 \clearlist{address}
 \clearfield{eprint}\clearfield{isbn}\clearfield{issn}
 \clearlist{location}\clearfield{month}\clearfield{series}
 \ifentrytype{book}{}{\clearlist{publisher}\clearname{editor}}
}
\newcommand\vldbdoi{XX.XX/XXX.XX}

\newcommand\vldbavailabilityurl{URL_TO_YOUR_ARTIFACTS}

\begin{document}


\title{Efficient Data Management with a Flexible Address Space}


\author{Chen Chen, Wenshao Zhong, and Xingbo Wu}
\affiliation{%
 \institution{\it University of Illinois at Chicago}
}
\email{{cchen262, wzhong20, wuxb}@uic.edu}


\renewcommand{\shortauthors}{Chen et al.}

\begin{abstract}

Data management applications store their data using structured files in which
data are usually sorted to serve indexing and queries.
However,
in-place insertions and removals of data are not naturally supported in a file's address space.
To avoid repeatedly rewriting existing data in a sorted file to admit changes in place,
applications usually employ extra layers of indirections, such as mapping tables and logs,
to admit changes out of place.
However, this approach leads to increased access cost and excessive complexity.

This paper presents a novel storage engine that provides a \textit{flexible address space},
where in-place updates 
of arbitrary-sized data, such as insertions and removals, can be performed efficiently.
With this mechanism, applications can
manage sorted data in a linear address space with minimal complexity.
Extensive evaluations show that a key-value store built
on top of it can achieve high performance and efficiency with a simple implementation.

\end{abstract}

\settopmatter{printfolios=true} 
\maketitle







\section{Introduction}
\label{sec:introduction}

Data management applications
store data in files for persistent storage.
The data are usually sorted in a specific order so that
they can be correctly and efficiently retrieved in the future.
However, it is not trivial to make updates such as insertions and deletions in these files.
To commit in-place updates in a sorted file,
existing data may need to be rewritten to maintain the file's layout.
For example,
key-value (KV) stores such as LevelDB~\cite{leveldb} and RocksDB~\cite{rocksdb}
need to merge and sort KV pairs in their data files periodically,
causing repeated rewriting
of existing KV data~\cite{raju2017pebblesdb,lu2016wisckey,stratos2020kv}.

It has been conventional wisdom to rewrite data to keep data sorted and
gain a better access locality.
By co-locating logically adjacent data in the storage device,
the data can be quickly accessed in the future
with a minimal number of I/O requests,
which is crucial for traditional
storage technologies such as HDDs.
However, when managing data with new storage technologies that
provide more balanced random and sequential performance
(e.g., Intel's Optane SSDs~\cite{optane}),
access locality is less of a dominant factor of I/O performance~\cite{wu2019optane}.
In this scenario, data rewriting becomes less beneficial for
future accesses but still consumes enormous CPU and I/O
resources~\cite{lepers2019kvell,papagiannis2016tucana}.
Therefore, it may not be cost-effective
to rewrite data on these devices in exchange for a better locality.
Despite this, data management applications still need to keep
their data logically sorted for efficient access.
An intuitive solution is to relocate data
in the address space logically without physically rewriting them.
However, this is barely feasible 
because of the lack of support for logically relocating data
in a file's address space.

In practice, applications pay a high cost to keep data sorted by using extra indirections.
For example, using a B$^+$-Tree to index data needs to rewrite tree nodes on updates.
LSM-Trees rewrite data less aggressively by using a multi-level layout,
which slows down reads due to sort-merging data on the fly.
Additionally, committing changes to these structures requires extra mechanisms
such as barriers and flushes,
which inflates the cost of maintaining crash consistency,
leading to problems like redundant journaling~\cite{shen2014journaling, yang2014stack}.
If the storage layer can provide support for keeping data logically sorted,
applications can delegate the data organizing jobs to the storage layer,
instead of employing extra persistent indirections at the application level.
To achieve this goal, the storage layer can provide a \textit{flexible address space}
that supports in-place data insertions and removals,
so that the data can be easily sorted.

Much effort has been made towards this direction.
For example, a few popular file systems---Ext4, XFS, and F2FS---have
provided \textit{insert-range} and \textit{collapse-range} features for inserting or removing
a range of data
in a file's address space to support various
types of applications~\cite{lwninsertinghole, fallocate}.
However, these mechanisms have not been able to help applications
because of a few fundamental limitations.
First of all, they have rigid block-alignment requirements.
For example, inserting a record of only a few bytes to a sorted data file using the
\textit{insert-range} operation is not allowed.
Second, \textit{shifting} a range of address mappings is very
inefficient with conventional address space indexes.
Inserting a new (aligned) data segment to a file
needs to shift all the existing address mappings after the insertion point to make room for the new data.
The shift operation has $O(N)$ cost
($N$ is the number of extents or blocks in the file),
which can be very costly due to
intensive metadata updates and journaling.
Third, commonly used data indexing mechanisms cannot keep track of shifted 
contents in a address space.
For example, indexes using offsets to record data positions are no longer usable
because the offsets can be easily changed by shift operations.
Therefore, a co-design of applications and the storage layer
is necessary to realize the benefits of managing data in a flexible address space.

This paper introduces FlexSpace, a storage engine that
provides a \textit{persistent flexible address space} for data management applications.
The core of FlexSpace is an address space indexing structure, named FlexTree, that is derived from the B$^+$-Tree structure.
In a FlexTree, it takes $O(\log{N})$ time to perform a
shift operation in the address space,
which is asymptotically faster than that of existing index data structures with $O(N)$ cost.
We implement FlexSpace as a user-space library.
It adopts log-structured space management for write efficiency and 
performs defragmentation based on data access locality for cost-effectiveness.
It also employs logical logging~\cite{rodeh2008btree, zhan2018fullpath}
to commit metadata updates at low cost.

We build FlexDB, a KV store that demonstrates how to implement
efficient data management applications with a flexible address space.
Based on the advanced features provided by FlexSpace,
FlexDB is able to maintain a fully sorted order of all KV pairs in a persistent address space
without employing complex indirections or rewriting data intensively.
In the meantime, it has a simple structure and a small codebase.
That being said, FlexDB is a fully functional KV store
that not only supports regular KV operations like
\texttt{PUT}, \texttt{GET}, \texttt{DELETE} and \texttt{SCAN}, but also
integrates efficient mechanisms to support caching,
concurrent access, and crash consistency.
Evaluation results show that FlexDB has substantially reduced the data rewriting overheads.
It achieves up to $16\times$ and $3.3\times$ speed-ups for read and write operations, respectively,
compared to two I/O-optimized KV stores, RocksDB and KVell.

This paper makes three major contributions.
First, we introduce an address space indexing structure,
namely FlexTree, that enables
efficient shift operations (\S\ref{sec:flextree}).
Second, we build FlexSpace to realize a persistent flexible address space,
in which data management applications can perform high-speed in-place data insertions and removals (\S\ref{sec:flexspace}).
Third, we use FlexDB to demonstrate a performant KV store
that can be easily built based on a flexible address space (\S\ref{sec:flexdb}).
Furthermore, we thoroughly evaluate the efficacy of a flexible
address space and its usage for data management (\S\ref{sec:evaluation}).

\section{Limitations of File Address Spaces}
\label{sec:background}

\begin{figure}[t]
  \centering
  \includegraphics[width=\columnwidth]{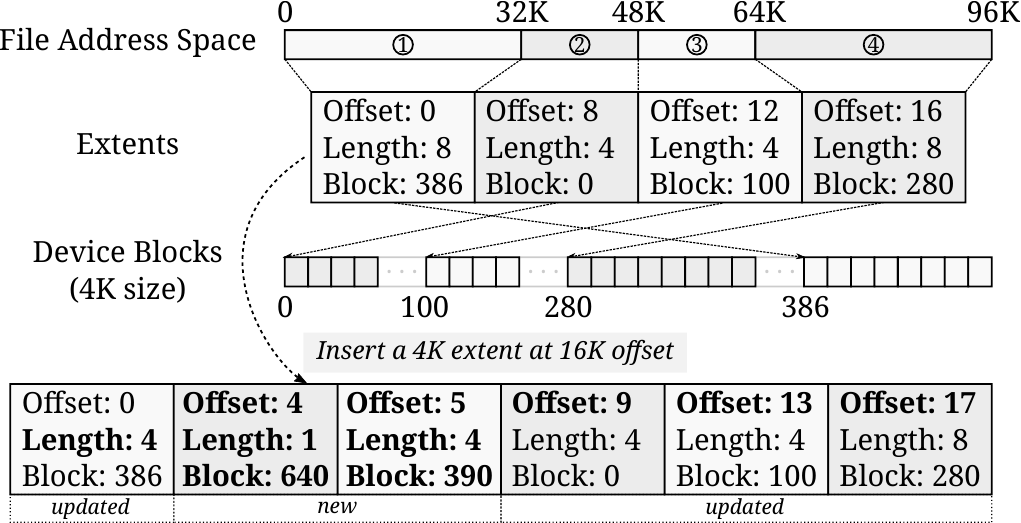}
  \caption{A file with a 96\,K address space}
  \label{fig:motiv}
\end{figure}

Modern file systems use extents to manage file address mappings.
An extent is a group of contiguous blocks. Its metadata
consists of three essential elements---file offset, length, and block number.
Figure~\ref{fig:motiv} shows an example of a
file with $96$\,KB address space on a file system using $4$\,KB blocks.
This file consists of four extents.
Real-world file systems employ index structures to manage extents.
For example, Ext4 uses an HTree~\cite{htree}.
Btrfs and XFS use a B$^+$-Tree~\cite{rodeh2013btrfs, xfsdoc}.
F2FS uses a multi-level mapping table~\cite{lee2015f2fs}.

Regular file operations such as overwrite do not modify existing mappings.
An append-write to a file needs to expand the last extent in place
or add new extents to the end of the mapping index,
which is of low cost.
However, the \textit{insert-range} and \textit{collapse-range}
operations in the aforementioned data structures
can be very expensive due to the shifting of extents.
To be specific, an \textit{insert-range} or \textit{collapse-range} operation
needs to update the offset value of
every extent after the insertion or removal point.
For example, inserting a 4\,KB extent at the offset of 16\,KB to the example file
in Figure~\ref{fig:motiv} needs to update all the existing extents' metadata.
Therefore, the shift operation has $O(N)$ cost, where $N$ is the total number of extents
after the insertion or removal point.

We benchmark the file editing performance
of an Ext4 file system on an Intel Optane 905P SSD.
There are three tests, namely, \textsc{pwrite}, \textsc{insert-range}, and \textsc{rewrite}.
\textsc{pwrite} starts with an empty file and uses the \texttt{pwrite} system call
to write 4\,KB blocks in a 1\,GB space in random order without overwrites.
Both \textsc{insert-range} and \textsc{rewrite} start with an empty file and
insert 4KB data blocks to random 4K-aligned offsets within the existing file space
until the file size reaches 1\,GB.
Accordingly, each insertion shifts the data after the insertion point forward.
\textsc{insert-range} utilizes the \textit{insert-range} operation of Ext4
(through the \texttt{fallocate} system call with \texttt{mode=FALLOC\_FL\_INSERT\_RANGE}) without rewriting file data.
\textsc{rewrite} rewrites the shifted data to realize insertion,
which on average rewrites half of the existing file for each insertion.

\begin{figure}[t!]
  \centering
  \includegraphics[width=0.8\columnwidth]{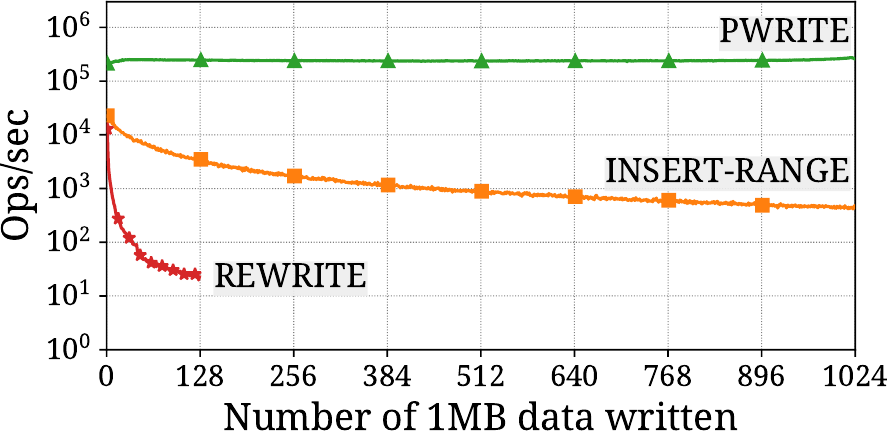}
  \caption{Performance of random write/insert on Ext4}
  \label{fig:fallocate}
\end{figure}

The results are shown in Figure~\ref{fig:fallocate}.
The \textsc{rewrite} test was terminated early due to its inferior performance
caused by intensive data rewriting.
Meanwhile, the test issued 5.5\,GB of write I/O to the SSD with only 128\,MB of new data inserted.
\textsc{insert-range} shows better performance than \textsc{rewrite}
by inserting data logically in the address space.
However, due to the inefficient shift operations in the extent index,
the throughput of \textsc{insert-range} dropped quickly and was eventually
nearly $1000\times$ lower than that of \textsc{pwrite}.
Although \textsc{insert-range} does not rewrite any user data,
it updates the metadata intensively and caused $25\%$ more writes to the SSD
compared to \textsc{pwrite}.
This number can be further increased if the application frequently calls
\texttt{fsync} to enforce write ordering.
XFS and F2FS also support the shift operations,
but they exhibit much worse performance than Ext4, so their results are not included.

Extents are simple and flexible for managing variable-length address mappings.
However, the alignment requirements
and the inefficient extent index structures in today's file address spaces hinder
the adoption of in-place data insertions and removals.
To make a flexible address space generally usable and affordable for data management applications,
an efficient mechanism that supports data shifting without rigid alignment requirements is indispensable.
\begin{figure*}[t!]
    \centering
    \begin{subfigure}[b]{0.45\textwidth}
        \centering
        \includegraphics[height=2.8cm]{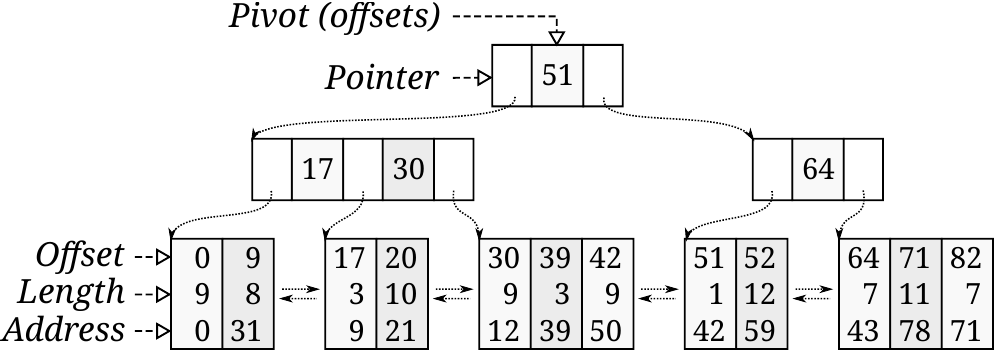}
        \caption{B$^+$-Tree}
        \label{fig:bptree}
    \end{subfigure}
    \quad
    \begin{subfigure}[b]{0.5\textwidth}
        \centering
        \includegraphics[height=2.8cm]{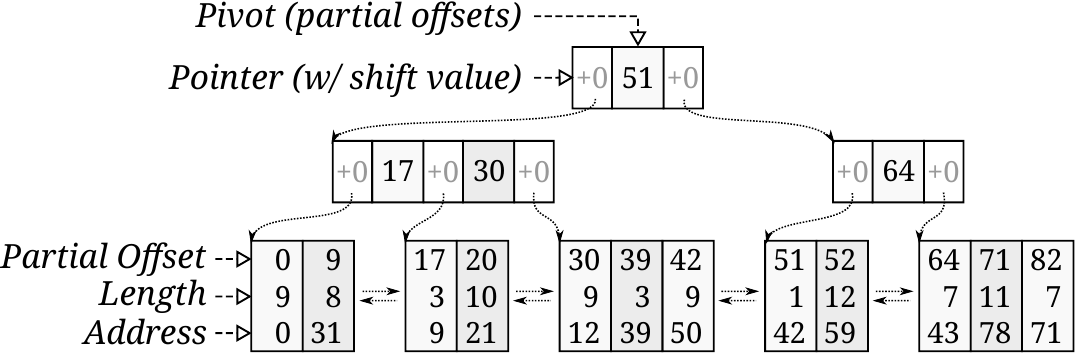}
        \caption{FlexTree}
        \label{fig:flextree}
    \end{subfigure}
    \caption{Examples of B$^+$-Tree and FlexTree that manage the same address space}
\label{figure:tree}
\end{figure*}

\section{FlexTree}
\label{sec:flextree}
Inserting or removing data in a file needs to
shift all the existing data beyond the insertion or removal point,
which causes intensive updates to the metadata of
the affected extents.
With regard to the number of extents in a file,
the cost of shift operations can be prohibitively
high due to the $O(N)$ complexity in existing extent index structures.

The following introduces FlexTree,
an augmented B$^+$-Tree that supports efficient shift operations.
The design of FlexTree is based on the observation that a shift operation
alters a contiguous range of extents.
FlexTree treats the shifted extents as a whole and
applies the updates to them collectively.
To facilitate this, it employs a new metadata representation scheme that
stores the address information of an extent on its search path.
As an extent index, it costs $O(\log{N})$ time to perform
a shift operation in FlexTree, and a shift operation only needs to update a few tree nodes.

\subsection{The Structure of FlexTree}
\label{sec:tree-struct}

Before demonstrating the design of FlexTree,
we first start with an example of B$^+$-Tree~\cite{algorithms}
that manages an address space
in byte granularity (Figure~\ref{fig:bptree}).
Each extent corresponds to a leaf-node entry consisting of
three elements---\textit{offset}, \textit{length}, and (physical) \textit{address}.
Each internal node contains \textit{pivot} entries
that separate the pointers to the child nodes.
When inserting a new extent at the head of an address space,
every existing extent's offset and every pivot's offset must be updated
because of the shift operation on the entire address space.

FlexTree employs an address metadata representation scheme
that allows for shifting extents with substantially reduced changes.
Figure~\ref{fig:flextree} shows a FlexTree
that encodes the same address mappings in the B$^+$-tree.
In FlexTree, the offset fields in extent entries and
pivot entries are replaced by \textit{partial offset} fields.
Besides, the only structural difference is that in a FlexTree,
every pointer to a child node is associated with a \textit{shift} value.
These shift values are used for encoding address information
in cooperation with the partial offsets.
The effective offset of an extent or pivot entry is determined by
the sum of the entry's partial offset and the shift values of the pointers
found on the search path from the root node to the entry.
The search path from the root node (at level 0) to an entry at level $N$
can be represented by a sequence
$\big((X_0, S_0), (X_1, S_1), \dots, (X_{N-1}, S_{N-1})\big)$,
where $X_i$ represents the index of the pointer at level $i$,
and $S_i$ represents the shift value associated with that pointer.
Suppose the partial offset of an entry is $P$.
Its effective offset $E$ can be calculated as $E=\big(\sum_{i=0}^{N-1}S_i\big)+P$.

\subsection{FlexTree Operations}

FlexTree supports basic extent operations such as
appending extents at the end of an address space and remapping existing extents,
as well as advanced operations,
including inserting or removing extents in the middle of an address space 
(\textit{insert-range} and \textit{collapse-range}).
The following explains how the address range operations execute in a FlexTree.
In this section, a leaf node entry in FlexTree is denoted by a triple:
\textit{(partial\_offset, length, address)}.

\paragraph{The insert-range Operation}
\label{subsec:insertion}

Inserting a new extent of length $L$ to a leaf node $z$ in FlexTree takes three steps.
First, the operation searches for the leaf node and inserts a new entry with a partial offset
$P=E-\big(\sum_{i=0}^{N-1}S_i\big)$, assuming the leaf node is not full.
When inserting to the middle of an existing extent, the extent must be split before the insertion.
The insertion requires a shift operation on all the extents after the new extent.
In the second step, for each extent within node $z$ that needs shifting,
its partial offset is incremented by $L$.
The remaining extents that need shifting span all the leaf nodes after node $z$.
We observe that, if every extent within a subtree needs to be shifted,
the shift value can be recorded in the pointer that points to the root of the subtree.
Therefore, in the third step, the remaining extents are shifted as a whole
by updating a minimum number of pointers
to a few subtrees that cover the entire range.
To this end, for each ancestor node of $z$ at level $i$,
the shift values of the pointers and the partial offsets of the pivots
after the pointer at $X_i$ are all added by $L$.
In this process, the updated pointers cover all the remaining extents,
and the path of each remaining extent contains exactly one updated pointer.
When the update is finished,
every shifted extent has its effective offset added by $L$.
The number of updated nodes of a shift operation is bounded by the tree's height,
so the operation's cost is $O(\log{}N)$.

\begin{figure}[t]
  \centering
  \includegraphics[width=0.95\columnwidth]{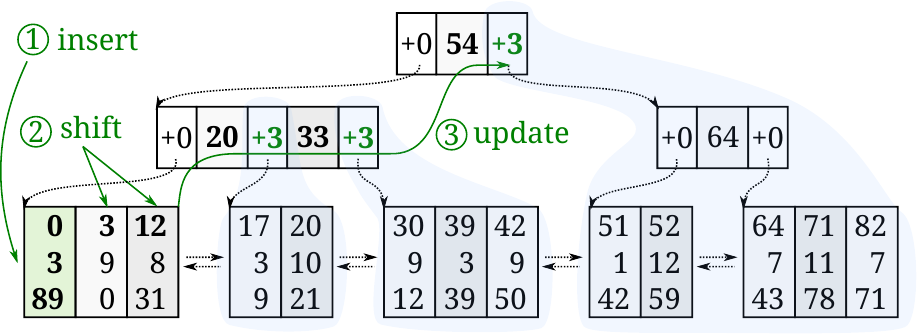}
  \caption{Inserting a new extent in FlexTree}
  \label{fig:ftinsert}
\end{figure}

Figure~\ref{fig:ftinsert} shows the process of inserting
a new extent with length $3$ and physical address $89$ to offset $0$ in the
FlexTree shown in Figure~\ref{fig:flextree}.
The first step is to search for the target leaf node for insertion.
Because all the shift values of the pointers are $0$,
the effective offset of every entry is equal to its partial offset.
Therefore, the target leaf node is the leftmost one,
and the new extent should be inserted at the beginning of that leaf node.
Then, there are three changes to be made to the FlexTree.
First, a new entry $(0,3,89)$ is inserted at the beginning of the target leaf node.
Second, the other two extents in the same leaf node are updated from
$(0,9,0)$ and $(9,8,31)$ to $(3,9,0)$ and $(12,8,31)$, respectively.
Third, following the target leaf node's path upward,
the pointers to the three subtrees covering the remaining leaf nodes and the corresponding pivots are updated,
as shown in the shaded areas in Figure~\ref{fig:ftinsert}.
Now, the effective offset of every existing leaf entry is increased by $3$.

\begin{figure}[ht]
  \centering
  \includegraphics[width=0.8\columnwidth]{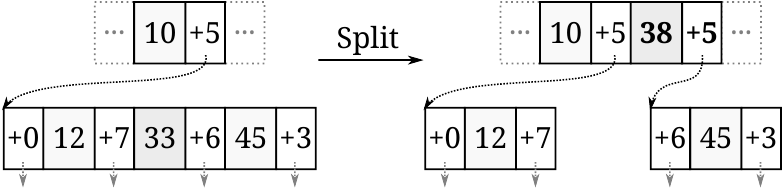}
  \caption{An example of node splitting in FlexTree}
  \label{fig:split}
\end{figure}

FlexTree splits every full node when a search travels down the tree for insertion.
The split threshold in FlexTree is one entry smaller than the node's capacity because
an insertion may cause an extent to be split,
which leads to two entries being added to the node for the insertion.
To split a node, half the entries in the node are moved to a new node.
Meanwhile, a pointer to the new node and a new pivot entry is created at the parent node.
The new pointer inherits the shift value of the pointer to the old node
so that the effective offsets of the moved entries remain unchanged.
The new pivot entry inherits the effective offset of the median key in the old full node.
The partial offset
of the new pivot is calculated
as the sum of the old median key's partial offset and the new pointer's inherited shift value.
Figure~\ref{fig:split} shows an example of a split operation.
The new pivot's partial offset is 38 (which is $5 + 33$).

\paragraph{Querying Mappings in an Address Range}

To retrieve the mappings of an address range in FlexTree,
the operation first searches for the starting point of the range,
which is a byte address within an extent.
Then, it scans forward on the leaf level from the starting point
to retrieve all the mappings in the requested range.
The correctness of the forward scanning is guaranteed by
the assumption that all extents on the leaf level are contiguous in the logical address space.
Apparently, a hole (an unmapped address range) in the logical address space can
break the continuity and lead to incorrect range size calculation and wrong search results.
To address this issue,
FlexTree explicitly records holes as unmapped ranges using entries with a special address value.

\begin{figure}[t]
  \centering
  \includegraphics[width=0.95\columnwidth]{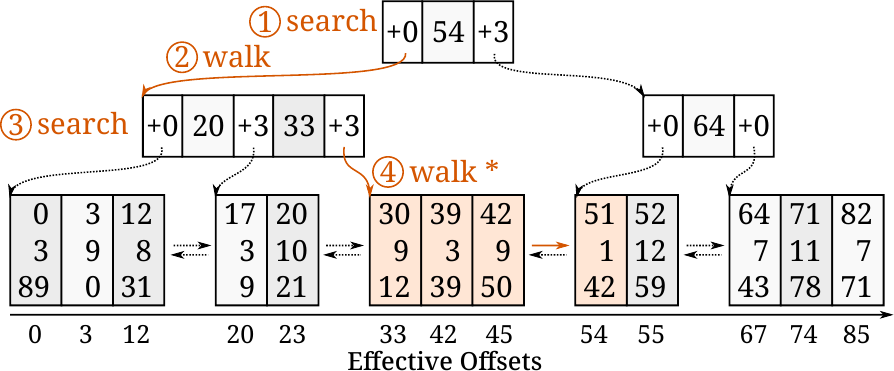}
  \caption{Looking up mappings from $36$ to $55$ in FlexTree}
  \label{fig:ftlookup}
\end{figure}

Figure~\ref{fig:ftlookup} shows the process
of querying the address mappings from $36$ to $55$, a $19$-byte range
in the FlexTree after the insertion in Figure~\ref{fig:ftinsert}.
First, a search of logical offset $36$ identifies the third leaf node.
The partial offset values of the pivots in the internal
nodes on the path are equal to their effective offsets ($54$ and $33$),
and the target leaf node has the path $\big((0, +0), (2, +3)\big)$.
Although the first extent in the target leaf node has a partial offset value of $30$,
its effective offset is $33$ ($0+3+30$).
Therefore, the starting point (logical offset $36$) is
the fourth byte within the first extent in the leaf node.
Then the address mappings of the 19-byte range can be retrieved by scanning the leaf nodes from that point.
The result is $\big((15, 6), (39, 3), (50, 9), (42, 1)\big)$,
an array of four tuples, each containing a physical address and a length.

\paragraph{The collapse-range Operation}

To collapse (remove without leaving a hole) an address range in FlexTree,
the operation first searches for the starting point of the removal.
If the starting point is in the middle of an extent, the extent is split
so that the removal will start from the beginning of an extent.
Similarly, a split is also used when the ending point is in the middle of an extent.
The address range being removed will cover one or multiple extents.
For each extent in the range, the extents after it are shifted backward
using a process similar to the forward shifting in the insertion operation
(\S\ref{subsec:insertion}).
The only difference is that a negative shift value is used.

\begin{figure}[h]
  \centering
  \includegraphics[width=0.96\columnwidth]{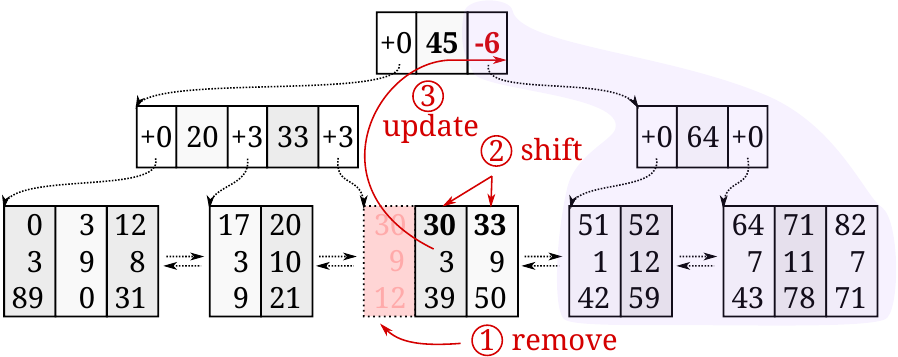}
  \caption{Removing address mapping from offset $33$ to $42$}
  \label{fig:ftdelete}
\end{figure}

Figure \ref{fig:ftdelete} shows the process of removing
a 9-byte address range ($33$ to $42$) from the FlexTree in Figure~\ref{fig:ftlookup}
without leaving a hole in the address space.
First, a search identifies the starting point,
which is the beginning of the first extent $(30 ,9 ,12)$ in the third leaf node.
Then the extent is removed, and the remaining extents in the leaf node are shifted backward.
Finally, in the root node, the pointer to the subtree that covers the last two leaf nodes
is updated with a negative shift value of $-9$,
as shown in the shaded area in Figure~\ref{fig:ftdelete}.

\begin{figure}[ht]
  \centering
  \includegraphics[width=0.85\columnwidth]{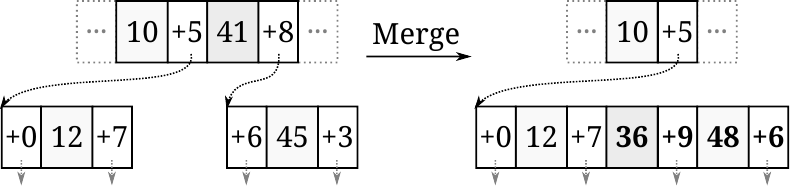}
  \caption{An example of node merging in FlexTree}
  \label{fig:merge}
\end{figure}

FlexTree merges a node to a sibling if their
total size is under a threshold after a removal.
Since two nodes being merged can have different shift values in their parents' pointers,
we need to adjust the partial offsets in the merged node to maintain correct effective offsets for all the entries.
When merging two internal nodes, the shift values are also adjusted accordingly.
Figure~\ref{fig:merge} shows an example of merging two internal nodes.

\subsection{Implementation}
\label{sec:flextree-imp}

FlexTree manages extent address mappings in byte granularity.
To be specific, the size of an extent can be an arbitrary number of bytes.
In the implementation of FlexTree, the internal nodes have 64-bit shift values and pivots.
For leaf nodes, we use 32-bit lengths, 48-bit partial offsets, and 48-bit physical addresses for extents.
The largest physical address value ($2^{48}-1$) is reserved for unmapped address ranges.

An effective offset can address a 64-bit space using the sum of 64-bit shift values and a 48-bit partial offset.
When a leaf node's maximum partial offset becomes too large,
to avoid overflow, FlexTree subtracts a value $M$, which is the minimum partial offset in the node,
from every partial offset of the node,
and adds $M$ to the node's corresponding shift value in the parent node.
Within a leaf node, the extents can cover up to 256\,TB, which is sufficiently large in practice.
\section{FlexSpace}
\label{sec:flexspace}

FlexSpace is a storage engine that provides persistent data storage in a flexible address space.
With FlexSpace, applications can make a better tradeoff by leveraging the lightweight in-place insertion/removal operations
to manage sorted data without using extra indirections or repeated data rewriting.

We implement FlexSpace as a user-space library.
It supports common file operations such as
\texttt{read}, \texttt{write}, \texttt{pread}, and \texttt{pwrite}.
It also provides advanced \texttt{insert\_range} and \texttt{collapse\_range} APIs
for in-place data insertions and removals.
The library enables concurrent access to individual address spaces using reader-writer locks.
It does not employ automated readahead since I/O efficiency
is often better exploited from the application level~\cite{lepers2019kvell,kannan2018novelsm,kourtis2019udepot}.

Internally, a FlexSpace's data and metadata are stored in regular files in a traditional file system.
Each FlexSpace consists of three files---a data file, a FlexTree file, and a logical log file.
The user-space library implementation gives FlexSpace the flexibility
to perform byte-granularity space management without any block alignment limitations.
In the meantime, the FlexSpace library delegates the job of cache management
to the operating system.

\subsection{Space Management}
\label{sec:flexspace_space}

A FlexSpace stores its data in a data file.
The data file's space is divided into fixed-size segments (4\,MB in the implementation),
which is similar to the structures in log-structured
storage systems~\cite{rosenblum1992lfs,lee2015f2fs,rumble2014log}.
Each new extent is allocated within a segment.
Specifically, a large write operation may create multiple logically contiguous extents
residing in different segments.
To avoid small writes,
an in-memory segment buffer is maintained, where consecutive extents
are automatically merged if they are logically contiguous.

The FlexSpace library performs garbage collection (GC) to reclaim space from underutilized segments.
It maintains an in-memory array to record the valid data size of each segment.
A GC process scans the array to identify a set of most underutilized segments
and relocates all the valid extents from these segments to new segments.
Then, the FlexTree extent index is updated accordingly.
Since the extents in a FlexSpace can have arbitrary sizes,
the GC process may produce less free space than expected because of the internal fragmentation in each segment.
To address this issue, we adopt an approach used
by a log-structured memory allocator~\cite{rumble2014log} to guarantee that
a GC process can always make forward progress.

By limiting the maximum extent size to $\frac{1}{K}$ of the segment size,
relocating extents in one segment whose utilization ratio is not higher than $\frac{K-1}{K}$
can reclaim free space for at least one new extent.
Therefore, if the space utilization ratio of the data file is capped at $\frac{K-1}{K}$,
the GC can always reclaim space from the most underutilized segment for writing new extents.
In the implementation,
we set the maximum extent size to be $\frac{1}{32}$ (128\,KB) of the segment size and conservatively limit the space
utilization ratio of the data file to $\frac{30}{32}$ ($93.75\%$).
In addition, we reserve at least $64$ free segments for relocating extents in batches.
The FlexSpace library also provides a \texttt{flexspace\_defrag} interface for manually
relocating a range of data in the file into new segments.
We will evaluate the efficiency of the GC policy in \S\ref{sec:evaluation}.

\subsection{Persistency and Crash Consistency}

A FlexSpace maintains an in-memory FlexTree that periodically synchronizes
its updates to the FlexTree file.
It must ensure atomicity and crash consistency in this process.
An insertion or removal operation often updates multiple tree nodes along the search path in the FlexTree.
If we use a block-based journaling mechanism to commit updates,
every dirtied node in the FlexTree will be written twice.
To address the potential performance issue, we use a combination of Copy-on-Write (CoW)~\cite{rodeh2008btree}
and logical logging~\cite{rodeh2008btree,zhan2018fullpath} to minimize the synchronization and I/O cost.

\paragraph{CoW}
CoW is used to synchronize the persistent FlexTree with the in-memory FlexTree.
The FlexTree file has a header at the beginning of the file that
contains a version number and a root node position.
A commit to the FlexTree file creates a new version of the FlexTree in the file.
In the commit process,
dirtied nodes are written to free space in the FlexTree file without rewriting existing nodes.
Once all the updated nodes have been written,
the file's header is updated atomically to make the new version persist.
Once the new version has been committed,
the file space used by the updated nodes in the old version can be safely reused in future commits.

\paragraph{Logical Logging}
Updates to the FlexTree extent index can be intensive
with small insertions and removals.
If every metadata update on the FlexTree directly commits to the FlexTree file,
the I/O cost can be high because every commit can update multiple tree nodes in the FlexTree file.
The FlexSpace library adopts the logical logging mechanism~\cite{rodeh2008btree, zhan2018fullpath}
to further reduce the metadata I/O cost.
Instead of performing CoW to the persistent FlexTree on every metadata update,
the FlexSpace library records every FlexTree operation in a log file
and only synchronizes the FlexTree file with the in-memory FlexTree
when the logical log has accumulated a sufficient amount of updates.
A log entry for an insertion or removal operation contains
the logical offset, length, and physical address of the operation.
A log entry for a GC relocation contains
the old and new physical addresses and the length of the relocated extent.
Each log entry takes 24 bytes of space (including 2 bits for the operation type),
which is much smaller than the node size of FlexTree.
The logical log can be seen as a sequence of operations that
transforms the persistent FlexTree to the latest in-memory FlexTree.
The version number of the persistent FlexTree is recorded at the head of the log.
Upon a crash-restart, uncommitted updates to the persistent FlexTree can be recovered
by replaying the log on the persistent FlexTree.

\paragraph{Write Ordering}
When writing data to a FlexSpace,
the data are first written to free segments in the data file.
Then, the metadata updates are applied to the in-memory FlexTree
and recorded in an in-memory buffer of the logical log.
The buffered log entries are committed to the log file periodically or on-demand for persistence.
In particular, the buffered log entries are committed after every execution of the GC process
to make sure that the new positions of the relocated extents are persistently recorded.
Then, the reclaimed space can be safely reused.
Upon a commit to the log file, the data file must be first synchronized
so that the logged operations will refer to correct file data.
When the logical log file size reaches a pre-defined threshold,
or the FlexSpace is being closed,
the in-memory FlexTree is synchronized to the FlexTree file using the CoW mechanism.
Afterward, the log file can be truncated and reinitialized using the FlexTree's new version number.

\begin{figure}[th]
\centering
\includegraphics[width=\columnwidth]{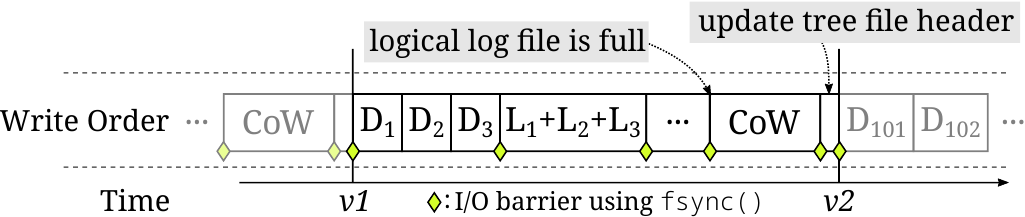}
\caption{An example of write ordering in FlexSpace}
\label{fig:ordering}
\end{figure}

Figure~\ref{fig:ordering} shows an example of the write ordering of a FlexSpace.
$D_i$ and $L_i$ represent the data write and the logical log write for the $i$-th file operation, respectively.
At the time of ``v1'', the persistent FlexTree (version 1) is identical to the in-memory FlexTree.
Meanwhile, the log file is almost empty, contains only a header that records the FlexTree version (version 1).
Then, for each write operation, the data is written to the data file (or buffered if the data is small), and
its corresponding metadata updates are logged in the logical log buffer.
When the logical log buffer is full, all the file data ($D_1$, $D_2$, and $D_3$)
are synchronized to the data file.
Then the buffered log entries ($L_1+L_2+L_3$) are written to the logical log file.
When the log file is full,
the current in-memory FlexTree is committed to the FlexTree file to create a new version (version 2)
in the FlexTree file using CoW.
Once the nodes have been written to the FlexTree file,
the new version number and the root node position of the FlexTree
are written to the file atomically.
The logical log is then cleared for recording future operations based on the new version.
I/O barriers (\texttt{fsync}) are used
before and after each logical log file commit and each FlexTree file header update
to enforce write ordering, as shown in Figure~\ref{fig:ordering}.
\section{FlexDB}
\label{sec:flexdb}

We build FlexDB, a KV store powered by
the advanced features of FlexSpace.
Just like the popular LSM-tree KV stores, LevelDB~\cite{leveldb} and RocksDB~\cite{rocksdb},
FlexDB buffers updates in a MemTable and
writes to a write-ahead log (WAL) for immediate data persistence.
When committing updates to the persistent storage,
however, FlexDB adopts a greatly simplified data model.
FlexDB stores all the KV pairs in sorted order in a FlexSpace
without using other persistent indirections.
Instead of performing repeated compactions across a multi-level store hierarchy
that causes high write amplification,
FlexDB directly commits updates from the MemTable to the FlexSpace in place at low cost.
FlexDB employs a space-efficient volatile sparse index to track positions of persistent KV data
in the FlexSpace and implements user-space caching for fast reads.

\subsection{Managing KV Data in a FlexSpace}
\label{sec:db-space}

FlexDB stores persistent KV pairs in a FlexSpace and keeps them always sorted
(in lexical order by default) with in-place updates.
Each KV pair in the FlexSpace starts with the key and value lengths
encoded with Base-128 Varint~\cite{dwarf128}, followed by the key and value's raw data.
A sparse KV index, whose structure is similar to a B$^+$-Tree,
is maintained in the memory to enable fast search in the FlexSpace.

\begin{figure}[t]
\centering
\includegraphics[width=0.99\columnwidth]{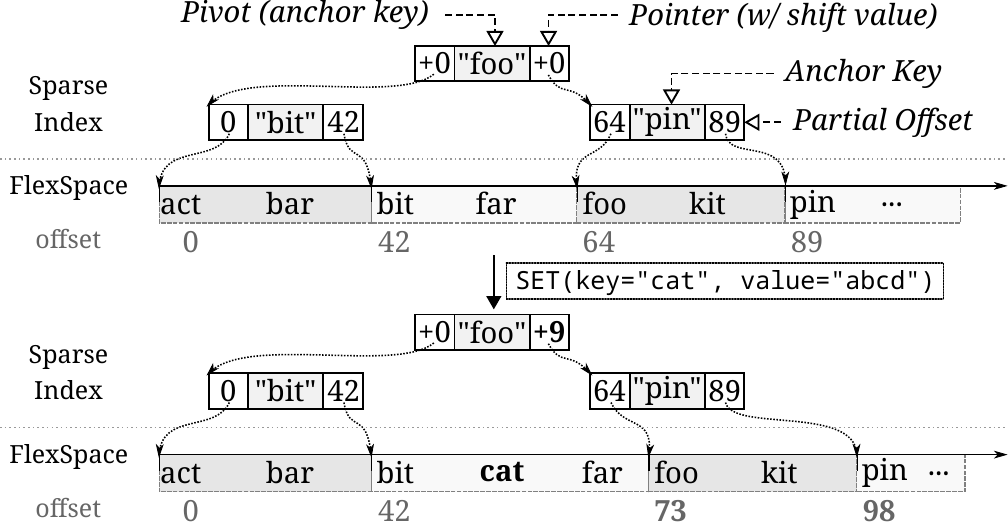}
\caption{An example of the sparse KV index in FlexDB}
\label{fig:flexdb}
\end{figure}

KV pairs in the FlexSpace are grouped into \textit{intervals},
each covering a number of consecutive KV pairs.
The sparse index stores an entry for each interval using
the smallest key in it as the index key.
The entry also records the size of the interval.
As with FlexTree, the sparse index encodes the offset of an interval
using the partial offset and the shift values on its search path.
Specifically, each leaf node entry contains a \textit{partial offset},
and each child pointer in internal nodes records a \textit{shift} value.
The effective offset of an interval is the sum of its partial offset and 
the shift values on its search path.
A search of a key performs a binary search
on the sparse index and calculates the effective offset of the interval.
Then, the search scans the interval to find the KV pair.
Figure~\ref{fig:flexdb} shows an example of the sparse KV index with four intervals.
The first interval does not need an index key.
The index keys of the other three intervals are ``bit'', ``foo'' and ``pin'', respectively.
A search of ``kit'' reaches the third interval (``foo'' $<$ ``kit'' $<$ ``pin'')
at offset 64 (0+64).

When inserting (or removing) a KV pair in an interval,
the offsets of all the intervals after it need to be shifted so that
the index can stay in sync with the FlexSpace.
The shift operation is similar to that in a FlexTree.
First, the operation updates the partial offsets
of the intervals in the same leaf node.
Then, the shift values on the path to the target leaf node are updated.
Different from that of FlexTree, the partial offsets in the sparse index are not the search keys
but the values in leaf node entries.
Therefore, the shift operation does not modify any index keys or pivots.
An update operation that resizes a KV pair is performed by
removing the old KV pair and inserting the new one at the same offset.

To insert a new KV pair (``cat'', ``abcd'') in the FlexDB shown in Figure~\ref{fig:flexdb}, a search first identifies the interval
at offset 42 whose index key is ``bit''.
Assuming the new KV item's size is 9 bytes,
we insert it to the FlexSpace between keys ``bit'' and ``far''
and shift the intervals after it forward by 9.
As shown at the bottom of Figure~\ref{fig:flexdb},
the effective offsets of the last two intervals are incremented by 9.

The sparse index needs to split a large interval or
merge two small intervals when their sizes reach specific thresholds.
The thresholds are specified by the total data size in bytes
and the number of KV pairs.
In the implementation, the split threshold is defined as
16\,KB and 16 KV items, whichever is exceeded first.
Two intervals can be merged if the total size is less than 16\,KB and they contain less than 16 KV items.

\subsection{Interval Caching}
\label{sec:db-cache}

Real-world workloads often exhibit skewed access
patterns~\cite{atikoglu2012workload, cao2020workload, yang2020twitter}.
Many popular KV stores employ user-space caching to exploit the access locality
for improved search efficiency~\cite{rocksdb,gilad2020evendb,conway2020splinterdb}.
FlexDB adopts a similar approach by caching frequently used intervals in the main memory.
The cache (namely the \textit{interval cache}) uses the CLOCK replacement algorithm~\cite{clock68} and a write-through policy.
Every interval's entry in the sparse index contains a cache pointer that is initialized as \texttt{NULL} to represent an uncached interval.
Upon a cache miss, a new cache entry is allocated by
creating an array of KV pairs based on the interval's data loaded from the FlexSpace.

When an interval is being loaded into the cache,
FlexDB marks it as \textit{fragmented}
if the number of extents is more than half the number of KV pairs.
When a marked interval is updated,
FlexDB uses \texttt{flexspace\_defrag}~(\S\ref{sec:flexspace_space})
to perform defragmentation on it.
In a cached interval, each KV pair is associated with a 16-bit hash fingerprint of the key for fast point queries
with a minimal number of key comparisons.
In range queries, a \texttt{SEEK} performs a binary search on the array.

\subsection{Supporting Concurrent Access}
\label{sec:db-co}

Updates in FlexDB are buffered in a MemTable.
The MemTable is a thread-safe skip list that supports
concurrent access of one writer and multiple readers.
Updates in the MemTable are periodically (or immediately when the MemTable is full)
committed to the FlexSpace and the sparse index by a background committer thread.
During this process, the MemTable becomes immutable and a new MemTable is created
to receive updates.
The committer can rewrite a highly fragmented interval for defragmentation
if the thread is not fully loaded.
A lookup in FlexDB first searches the MemTables.
If the key is not found,
it queries the sparse KV index to find the key in the FlexSpace.

When the committer thread is active, it requires exclusive access to the sparse index
and the FlexSpace to prevent inconsistent data or metadata from being reached by readers.
To this end, a reader-writer lock is used for the committer thread to block the readers 
when necessary.
For balanced performance and responsiveness, the committer thread releases and reacquires
the lock every 1000 KV pairs committed.
Therefore, readers can be served quickly
without waiting for the completion of the committing process.
We will measure and discuss the wait time in \S\ref{sec:eval_flexdb}.

\subsection{Crash Recovery and Index Rebuilding}
\label{sec:db-recovery}
Upon a restart, FlexDB first recovers the uncommitted KV data from the write-ahead log.
Then, it constructs the volatile sparse KV index.
Intuitively the sparse index can be built by
sequentially scanning the KV pairs in the FlexSpace,
but the cost can be significant in a large store.
In fact, the rebuilding only requires an index key for each interval.
Therefore, a sparse index could be quickly constructed by skipping
a certain amount of data every time an index key is determined.

In FlexDB, the FlexSpace's extents are created by
inserting or removing KV pairs,
which guarantees that every extent always begins with a KV pair.
To identify a KV pair in the middle of the FlexSpace without knowing its exact offset,
we add a \texttt{read\_extent(off, buf, maxlen)} function to the FlexSpace library.
The function searches for the extent at the designated offset (\texttt{off})
and reads up to \texttt{maxlen} bytes of data from the beginning of the extent.
The extent's size, logical offset ($\le$ \texttt{off}), and the number of bytes read are returned.
To build a sparse index,
\texttt{read\_extent} is used to retrieve a key at each approximate interval offset
(8\,KB, 16\,KB, \dots) and these keys are used as index keys of the new intervals.
FlexDB can immediately start processing requests once the sparse index is built.
A recovered interval whose size exceeds the split threshold will be split when it is accessed.
\section{Evaluation}
\label{sec:evaluation}

In this section, we experimentally evaluate FlexTree, the FlexSpace library, and FlexDB.
All the experiments are run on a server with
an Intel 10-core Xeon Silver 4210 CPU and 64\,GB RAM.
The persistent storage device of all tests
is an Intel Optane 905P SSD with 960\,GB capacity.
The workstation runs a 64-bit Linux OS with kernel version 5.10.32 LTS.

\subsection{FlexTree as Address Space Index}

First of all, we evaluate the performance
of the FlexTree index structure and compare it with a regular B$^+$-Tree and a sorted array.
In the evaluation of FlexTree, we want to answer the following questions:
(1) What is the practical performance advantage of the asymptotic $O(\log{N})$ shift operations
in FlexTree compared to data structures that have $O(N)$ cost?
(2) Can FlexTree efficiently handle range query,
which is frequently used for retrieving the address mapping information of a range of data?
(3) How much overhead does FlexTree introduce to common address space operations such as lookup and append,
compared to a regular B$^+$-Tree?

The B$^+$-Tree has the structure shown in Figure~\ref{fig:bptree},
which is identical to FlexTree except that the shift values are removed from the internal nodes.
In a shift operation,
the B$^+$-Tree and the array must update all the shifted extents.

We benchmark four index
operations---\textit{insert}, \textit{append}, \textit{lookup} and \textit{range-query}.
An insert experiment starts with an empty index.
Each operation inserts a new extent at a random offset within the existing space.
An append experiment starts with an empty index.
Each operation appends a new extent after the existing extents.
A lookup experiment randomly queries extents,
and every operation must search the index.
A range-query experiment randomly queries
ranges consisting of 50 extents,
where each operation searches for the first extent, then walks
on the leaf nodes or the array to read the next 50 extents.
These extent index structures are memory-resident and there are no persistent data.

Table~\ref{tab:flextree-test} shows the throughput of each data structure in the experiments.
Since FlexTree's address metadata representation scheme allows for much faster extent insertions,
it shows high throughput in the insert experiments.
However, the B$^+$-Tree and the sorted array
show extremely high overheads due to the intensive memory writes and movements.
To be specific, every time an extent is inserted at the beginning,
the entire mapping index is rewritten.
FlexTree maintains a consistent $O(\log{}N)$ cost for insertions,
which is asymptotically and practically faster.

\setlength{\tabcolsep}{2pt}
\begin{table}[t]
\centering
\caption{Throughput of the extent metadata operations}
\begin{tabular}{c|c||c|c||c|c|c|c|c|c}
\hline
\multicolumn{2}{c||}{Experiment} &  \multicolumn{2}{c||}{Insert} & \multicolumn{2}{c|}{Append} & \multicolumn{2}{c|}{Lookup} & 
\multicolumn{2}{c}{Range} \\ \hline

\multicolumn{2}{c||}{\rule{0pt}{10pt}\# of Extents}
& $10^5$ & $10^6$ & $10^8$ & $10^9$ & $10^8$ & $10^9$ & $10^8$ & $10^9$ \\ \hline
\parbox[t]{2.25mm}{\multirow{3}{*}{\rotatebox[origin=c]{90}{\footnotesize Mops/sec}}}
& FlexTree &     \multicolumn{1}{l|}{6.25}  & \multicolumn{1}{l||}{5.26}   & 14.56 & 13.30 & 11.7 & 9.17 & 7.40 & 6.37 \\ \cline{2-10}
& B$^+$-Tree &   0.026 & 0.0014 & 14.84 & 13.55 & 11.8 & 9.24 & 7.47 & 6.41 \\ \cline{2-10}
& Sorted Array & 0.028 & 0.0017 & 21.74 & 19.61 & 12.7 & 8.18 & 8.15 & 6.15 \\ \cline{2-10}
\end{tabular}
\label{tab:flextree-test}
\end{table}

For appends, the sorted array outperforms FlexTree and the B$^+$-Tree
because appending new extents at the end of an array
does not need node splits or memory allocations.
Meanwhile, FlexTree is only 2\% slower than the B$^+$-Tree.
In the lookup and range-query experiments, the sorted array also
outperforms the others when the number of extents is smaller ($10^8$).
However, with more extents ($10^9$), the array exhibits lower throughput than the other indexes.
This is because a binary search in the sorted array causes scattered memory accesses and
more cache misses than in tree nodes.
In the three experiments, the throughput of FlexTree and B$^+$-Tree are close,
which suggests that the calculation of effective offsets in FlexTree is of low cost.
FlexTree also inherits the good range query efficiency from B$^+$-Tree.

\setlength{\tabcolsep}{1.12pt}
\begin{table*}[t]
\centering
\caption{Single-threaded I/O performance of FlexSpace and regular files in XFS, Ext4, F2FS, and BtrFS}
\begin{tabular}{c|c||c c c|c c c c c|c c c c c||c c c|c c c c c|c c c c c}
\hline
\multicolumn{2}{c||}{I/O Size}
& \multicolumn{13}{c||}{4\,KB\quad(File Size = 1\,GB)}
& \multicolumn{13}{c}{64\,KB\quad(File Size = 16\,GB)}
\\
\hline

\multicolumn{2}{c||}{Write Pattern} &
\multicolumn{3}{c|}{Rand. Insert} & \multicolumn{5}{c|}{Rand. Write} &
\multicolumn{5}{c||}{Seq. Write} &
\multicolumn{3}{c|}{Rand. Insert} & \multicolumn{5}{c|}{Rand. Write} &
\multicolumn{5}{c}{Seq. Write} \\
\hline
\multicolumn{2}{c||}{System} & Flex & XFS & Ext &
                               Flex & XFS & Ext & F2 & Btr &
                               Flex & XFS & Ext & F2 & Btr &
                               Flex & XFS & Ext &
                               Flex & XFS & Ext & F2 & Btr &
                               Flex & XFS & Ext & F2 & Btr \\
\hline
\multicolumn{2}{c||}{Write (GB/s)} &
\ColRowGB{0.62} & \myepsilon & \myepsilon & \ColRowGB{0.61} & 
\ColRowGB{0.57} & \ColRowGB{0.50} & \ColRowGB{0.61} & \ColRowGB{0.62} & 
\ColRowGB{0.62} & \ColRowGB{0.63} & \ColRowGB{0.60} & \ColRowGB{0.55} & 
\ColRowGB{0.62} & \ColRowGB{0.75} & \myepsilon & \ColRowGB{0.05} & 
\ColRowGB{0.76} & \ColRowGB{0.79} & \ColRowGB{0.77} & \ColRowGB{0.85} & 
\ColRowGB{0.64} & \ColRowGB{0.77} & \ColRowGB{0.82} & \ColRowGB{0.77} & 
\ColRowGB{0.72} & \ColRowGB{0.82}
\\ 
\multicolumn{2}{c||}{W. A. Ratio} &
\ColRowAmp{1.03} & \ColRowAmp{5.53} & \ColRowAmp{2.23} & \ColRowAmp{1.03} & 
\ColRowAmp{1.02} & \ColRowAmp{1.06} & \ColRowAmp{1.02} & \ColRowAmp{1.10} & 
\ColRowAmp{1.03} & \ColRowAmp{1.02} & \ColRowAmp{1.05} & \ColRowAmp{1.02} & 
\ColRowAmp{1.03} & \ColRowAmp{1.02} & \ColRowAmp{1.83} & \ColRowAmp{1.10} & 
\ColRowAmp{1.02} & \ColRowAmp{1.02} & \ColRowAmp{1.05} & \ColRowAmp{1.03} & 
\ColRowAmp{1.03} & \ColRowAmp{1.02} & \ColRowAmp{1.02} & \ColRowAmp{1.05} & 
\ColRowAmp{1.03} & \ColRowAmp{1.03}
\\ \hline

\parbox[t]{2.6mm}{\multirow{4}{*}{\rotatebox[origin=c]{90}{Read (GB/s)}}} 
& Seq. Cold &
\ColSeCo{0.39} & \ColSeCo{1.11} & \ColSeCo{0.95} & \ColSeCo{0.41} & 
\ColSeCo{1.97} & \ColSeCo{1.82} & \ColSeCo{1.81} & \ColSeCo{1.15} & 
\ColSeCo{1.92} & \ColSeCo{1.97} & \ColSeCo{1.83} & \ColSeCo{1.81} & 
\ColSeCo{1.66} & \ColSeCo{0.95} & \ColSeCo{1.93} & \ColSeCo{2.07} & 
\ColSeCo{0.95} & \ColSeCo{1.92} & \ColSeCo{2.05} & \ColSeCo{2.02} & 
\ColSeCo{1.63} & \ColSeCo{1.93} & \ColSeCo{2.05} & \ColSeCo{2.05} & 
\ColSeCo{2.02} & \ColSeCo{1.71}
\\
& Rand. Cold &
\ColRaCo{0.38} & \ColRaCo{0.38} & \ColRaCo{0.36} & \ColRaCo{0.38} & 
\ColRaCo{0.41} & \ColRaCo{0.37} & \ColRaCo{0.40} & \ColRaCo{0.24} & 
\ColRaCo{0.40} & \ColRaCo{0.41} & \ColRaCo{0.36} & \ColRaCo{0.40} & 
\ColRaCo{0.25} & \ColRaCo{0.94} & \ColRaCo{0.97} & \ColRaCo{0.99} & 
\ColRaCo{0.94} & \ColRaCo{0.95} & \ColRaCo{1.00} & \ColRaCo{0.98} & 
\ColRaCo{0.79} & \ColRaCo{0.95} & \ColRaCo{1.01} & \ColRaCo{0.98} & 
\ColRaCo{1.00} & \ColRaCo{0.81}
\\
& Seq. Warm &
\ColSeWa{3.44} & \ColSeWa{4.26} & \ColSeWa{4.48} & \ColSeWa{3.44} & 
\ColSeWa{4.39} & \ColSeWa{4.44} & \ColSeWa{4.43} & \ColSeWa{4.44} & 
\ColSeWa{4.33} & \ColSeWa{4.42} & \ColSeWa{4.44} & \ColSeWa{4.42} & 
\ColSeWa{4.38} & \ColSeWa{5.45} & \ColSeWa{5.76} & \ColSeWa{5.70} & 
\ColSeWa{5.43} & \ColSeWa{5.70} & \ColSeWa{5.71} & \ColSeWa{5.73} & 
\ColSeWa{5.74} & \ColSeWa{5.58} & \ColSeWa{5.71} & \ColSeWa{5.74} & 
\ColSeWa{5.76} & \ColSeWa{5.74}
\\
& Rand. Warm &
\ColRaWa{2.42} & \ColRaWa{3.23} & \ColRaWa{2.75} & \ColRaWa{2.40} & 
\ColRaWa{3.28} & \ColRaWa{3.39} & \ColRaWa{3.41} & \ColRaWa{3.38} & 
\ColRaWa{2.77} & \ColRaWa{3.28} & \ColRaWa{3.36} & \ColRaWa{3.42} & 
\ColRaWa{3.38} & \ColRaWa{5.21} & \ColRaWa{5.46} & \ColRaWa{4.43} & 
\ColRaWa{5.17} & \ColRaWa{5.43} & \ColRaWa{5.44} & \ColRaWa{5.40} & 
\ColRaWa{5.49} & \ColRaWa{5.19} & \ColRaWa{5.42} & \ColRaWa{5.47} & 
\ColRaWa{5.42} & \ColRaWa{5.46}
\\ \cline{2-28}

\end{tabular}
\label{tab:flexspace-test}
\end{table*}

\subsection{The FlexSpace Library}

In this section, we evaluate the efficiency of data I/O operations in the FlexSpace library.
Note that FlexSpace is a storage engine that provides a persistent flexible address space
for data management applications.
Although there are overlaps between FlexSpace and file system functionalities, FlexSpace
does not replace file systems on managing traditional files and directories.
Therefore,
file system benchmarks that require hierarchy directory structures do not apply to FlexSpace.
In this section, we focus on data I/O and shifting operations within a persistent
address space.

We compare FlexSpace with file address spaces provided by
four representative file systems,
Ext4~\cite{ext4}, XFS~\cite{xfs}, F2FS~\cite{lee2015f2fs}, and Btrfs~\cite{rodeh2013btrfs}.
Among them, Ext4, XFS, and F2FS support block-aligned shift operations.
The four file systems are formatted using \texttt{mkfs} with their default arguments.
FlexSpace stores its internal files on an XFS file system.

In the evaluation of FlexSpace, we want to answer the following questions:
(1) What is the performance benefit of FlexSpace's
\textit{insert-range} and \textit{collpase-range} operations?
(2) How do different access patterns affect the performance of FlexSpace?
(3) What are the performance implications of implementing a storage engine
in the user space?

Each experiment consists of a write phase and a read phase with one thread.
There are three write patterns for the
write phase---random insert (using \textit{insert-range}),
random write, and sequential write.
The first two patterns are the same as the \textsc{insert} and \textsc{pwrite}
in Section~\ref{sec:background}, respectively.
The sequential write pattern writes data blocks sequentially.
A write phase starts with an empty address space and writes or inserts data blocks using the respective pattern.
Finally, an I/O barrier (\texttt{fsync} in file systems) is issued to enforcing I/Os.
Note that an I/O barrier in FlexSpace consists of flushing
all buffered segment writes, using CoW to checkpoint the updated in-memory FlexTree,
and calling \texttt{fsync} on all its internal files.
After the write phase,
we measure the read performance with two patterns---sequential and random.
Each read operation reads a block of data from the address space.
The random pattern uses randomly shuffled offsets so that
it reads each data block in the address space exactly once.
For each read pattern, the kernel page cache is first cleared.
Then the program reads the entire address space twice,
once with a cold cache and once with the cache warmed up.
In the experiments, we adopt two I/O sizes---4\,KB and 64\,KB.
With each I/O size, we use the same number of blocks ($2^{18}$)
to construct the address space.
Therefore, the address space sizes are 1\,GB and 16\,GB, respectively.
Table~\ref{tab:flexspace-test} shows the experimental
results ($\varepsilon$ represents a value {\small$<$}$0.01$).
We also include the write amplification (WA) ratios of each experiment,
derived from the SMART data of the SSD.
The following discusses the key observations.

\paragraph{Insert}

FlexSpace's random insert throughput can be up to
180$\times$ higher than Ext4 ($620$\,MB/s vs. $3.36$\,MB/s)
and four orders of magnitude higher than XFS.
F2FS exhibits lower throughput than XFS so its results are omitted.
Throughout the insertion process,
FlexSpace can maintain high throughput while Ext4 and XFS
suffer extreme throughput degradations because of the growing extent index sizes that
lead to increasingly intensive metadata updates.

\paragraph{Write}
The random and sequential write throughput of FlexSpace is on par with the other systems.
FlexSpace commits writes to the data file (stored in XFS)
in the unit of segments, which enables batching and buffering in the user space.
Meanwhile, FlexSpace adopts the log-structured write in the data file,
which transforms random writes on the FlexSpace into sequential writes on the SSD.
As a result, random writes in FlexSpace can outperform XFS with the 4KB I/O size.

\paragraph{Write Amplification}
In the random and sequential write experiments,
all the systems show low WA ratios because
the metadata updates are not intensive.
However, in the random insert experiments,
Ext4 and XFS show very high WA ratios (up to 5.53) because each insert operation updates
half of the existing extents' metadata on average,
which leads to intensive computation and metadata I/O.
XFS and Ext4's WA ratios are lower with the I/O size increased (64\,KB)
since the amount of metadata updates remains the same.
That said, they still show low throughput because of the high computation cost.
In FlexSpace, the insert operations have a very low cost ($O(\log{}N)$) and
the logical logging can further reduce metadata write.
As a result, FlexSpace achieves fast inserts ($\geq$620\,MB/s) with constantly low WA ratios ($\leq 1.03$).

\paragraph{Read}

All the systems show similar read speed on address spaces constructed with sequential writes.
However, with random writes/inserts, FlexSpace generates a fragmented data file layout
which causes random read in the data file.
As a result, when reading sequentially with a cold cache,
FlexSpace shows $2.8\times$ to $4.8\times$ lower throughput than the file systems.
That said, all the systems show slow random read with a cold cache
since there is hardly any readahead in the kernel.

\begin{figure}[t]
\centering
\includegraphics[width=\columnwidth]{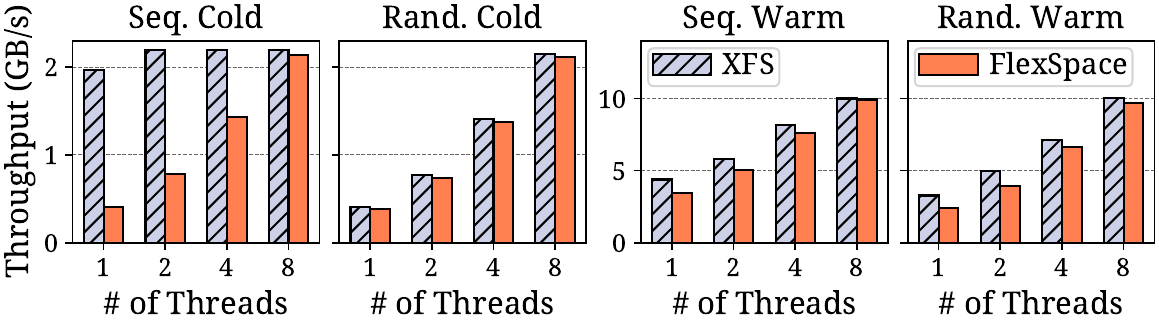}
\caption{Read throughput after random write (4\,KB)}
\label{fig:flexspace-multi}
\end{figure}

Data management systems often rely on asynchronous I/O or multi-threading to
exploit I/O bandwidth~\cite{lepers2019kvell,kannan2018novelsm,kourtis2019udepot}.
To evaluate the I/O efficiency in this context,
we run the read experiments with different numbers of threads.
As shown on the left of Figure~\ref{fig:flexspace-multi},
XFS's throughput is already near its peak with one thread because
of the automated readahead in the kernel.
FlexSpace's throughput continues to increase with more threads and
eventually reaches 98\% of XFS's throughput.

As shown in Figure~\ref{fig:flexspace-multi}, FlexSpace's throughput is
close to XFS when the cache is warmed up.
The difference is larger with fewer threads because of the constant costs
of accessing the FlexTree.
Like the previous experiment,
multi-threading can hide these costs and
also increase access throughput.
With eight threads, FlexSpace's throughput increased by up to $4\times$
and is at least 96\% of XFS's throughput.

\subsection{FlexDB Performance}
\label{sec:eval_flexdb}

\begin{figure*}[t]
    \centering
    \begin{subfigure}[b]{0.33\textwidth}
    \centering
        \includegraphics[height=3.2cm]{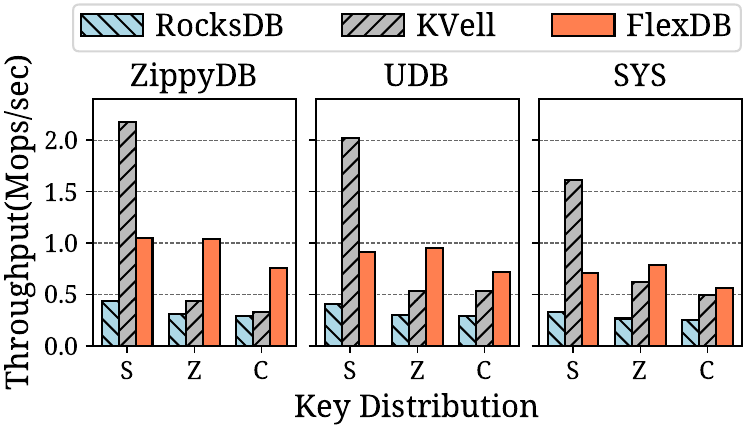}
        \caption{Insertion throughput}
        \label{fig:flexdb-micro-write}
    \end{subfigure}
    \hfill
    \begin{subfigure}[b]{0.33\textwidth}
    \centering
        \includegraphics[height=3.2cm]{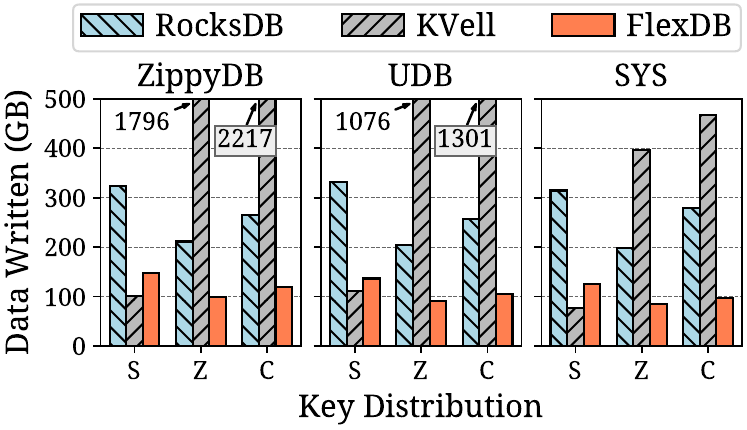}
        \caption{SSD writes}
        \label{fig:flexdb-micro-wa}
    \end{subfigure}
    \hfill
    \begin{subfigure}[b]{0.33\textwidth}
    \centering
        \includegraphics[height=3.2cm]{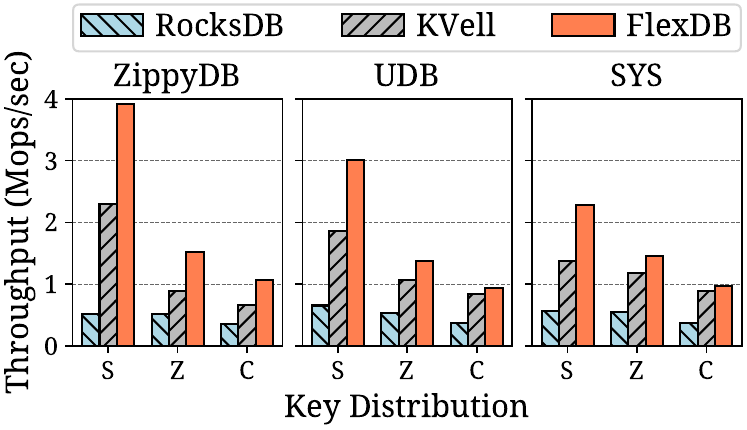}
        \caption{Point query throughput}
        \label{fig:flexdb-micro-read}
    \end{subfigure}
    \begin{subfigure}[b]{0.38\textwidth}
    \centering
        \includegraphics[height=3.2cm]{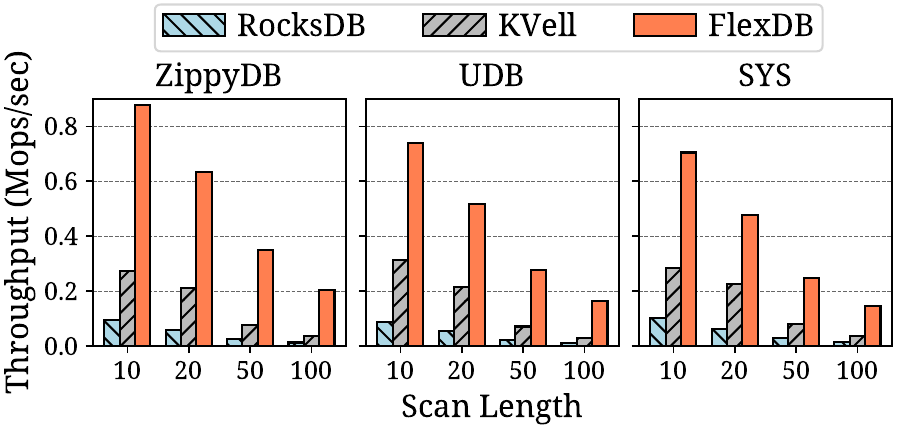}
        \caption{Scan throughput (Zipfian)}
        \label{fig:flexdb-micro-scan}
    \end{subfigure}
    \hfill
    \begin{subfigure}[b]{0.38\textwidth}
    \centering
        \includegraphics[height=3.2cm]{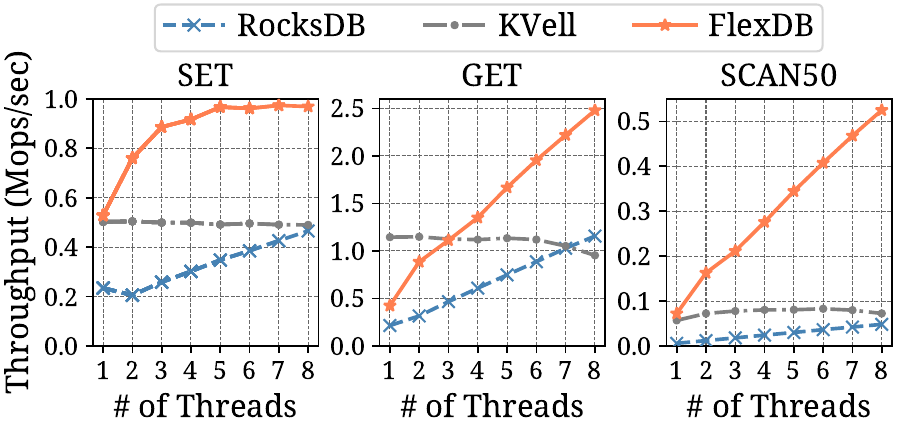}
        \caption{Scalability (UDB+Zipfian)}
        \label{fig:flexdb-scale}
    \end{subfigure}
    \hfill
    \begin{subfigure}[b]{0.23\textwidth}
    \centering
        \includegraphics[height=3.2cm]{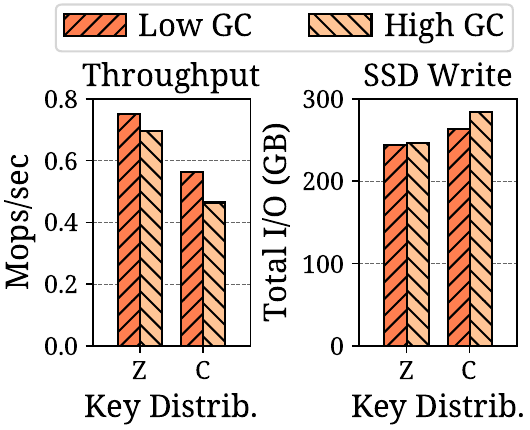}
        \caption{GC overhead}
        \label{fig:flexdb-gc}
    \end{subfigure}
    \caption{Microbenchmark results of FlexDB. Key distributions:
    \textit{S} -- Sequential; \textit{Z} -- Zipfian;
    \textit{C} -- Zipfian-Composite.}
\end{figure*}

The goal of FlexDB is to demonstrate that a simple persistent KV store
built based on a persistent flexible address space (FlexSpace) can match or outperform
the state-of-the-arts that are built based on traditional files.

We evaluate the performance of FlexDB through various experiments
and compare it with Facebook's RocksDB~\cite{rocksdb},
a representative LSM-Tree KV store, and KVell~\cite{lepers2019kvell},
an NVMe-optimized B$^+$-Tree-based KV store that
exploits I/O bandwidth with asynchronous I/O
and uses a full index in memory for fast search.
We also evaluated LMDB (B$^+$-Tree based)~\cite{lmdb}
and TokuDB (B$^\varepsilon$-Tree based)~\cite{tokudb}.
However, they exhibit consistently low performance compared with RocksDB.
Similar observations are also reported in recent studies~\cite{gilad2020evendb,papagiannis2016tucana,dong2017optimizing}.
Therefore, their results are omitted.

For a fair comparison, both FlexDB and RocksDB are configured
with 1\,GB MemTables and 16\,GB user-space cache.
RocksDB is tuned as suggested by its official tuning guide
(following the configurations for ``Total ordered database, flash storage.'')~\cite{rocksdb-tuning}.
FlexDB has its automatic defragmentation and the FlexSpace GC always enabled.
KVell maintains its own page cache in the user space and uses
direct I/O to bypass the kernel's cache.
We adjust its cache size ($\geq$16\,GB) based on the actual memory footprint in each experiment
to make sure it can fully utilize the available memory on the machine.
Compression is disabled in all the stores.

All the experiments in this section run with 4 concurrent client threads unless otherwise noted.
FlexDB uses only one background thread (the committer thread described in \S\ref{sec:db-co}).
RocksDB has up to 4 background compaction threads.
KVell is configured with 4 worker threads, each with an I/O depth of 64.
Therefore, the numbers of CPU cores that can be utilized by
FlexDB, RocksDB, and KVell are 5, 8, and 8, respectively.
For read and YCSB experiments,
each data point is measured by running the respective workload for 60s.

We generate synthetic KV datasets using the representative KV sizes of Facebook's production
workloads~\cite{atikoglu2012workload, cao2020workload}.
Table~\ref{tab:kvsize} shows the details of the datasets.
The size of each dataset is about 64\,GB, approximately
4$\times$ the size of the user-level cache in FlexDB and RocksDB.
The workloads are generated using three key distributions---sequential, Zipfian ($\alpha=0.99$),
and Zipfian-Composite~\cite{gilad2020evendb}.
With Zipfian-Composite,
the prefix (the first three decimal digits) of a key follows the default Zipfian distribution,
and the remaining are drawn uniformly at random.

\setlength{\tabcolsep}{5.2pt}
\begin{table}[t]
\centering
\caption{Synthetic KV datasets with real-world KV sizes}
\begin{tabular}{c||c|c|c}
\hline
Dataset & ZippyDB~\cite{cao2020workload} &
UDB~\cite{cao2020workload} & SYS~\cite{atikoglu2012workload}
\\ \hline
Avg. Key+Value Sizes (B) & 48+43 & 27+127  & 28+396 \\ \hline
Number of KV pairs & 720\,M & 420\,M & 150\,M \\
\hline FlexDB Index Size & 3.51\,GB & 1.50\,GB & 534\,MB \\
\hline
\end{tabular}
\label{tab:kvsize}
\end{table}

\paragraph{Write (PUT)}
Each write experiment starts from an empty store.
Each client thread inserts
$25\%$ (approx. 16\,GB) of the dataset
to the store following the key distribution.
For sequential load,
the dataset is partitioned into four contiguous ranges,
and each thread inserts one range of KV pairs.
For the Zipfian and Zipfian-Composite distributions,
existing keys can be overwritten,
which leads to reduced write I/O if Memtables are used.

Figures~\ref{fig:flexdb-micro-write} and \ref{fig:flexdb-micro-wa} show the
measured throughput and amount of disk I/O of the systems.
KVell outperforms FlexDB and RocksDB by more than 2$\times$
with sequential load, which is because
KVell fully utilizes the I/O bandwidth without writing to a WAL.
In comparison, FlexDB has only one committer thread and
needs to record KV pairs in the WAL.
Meanwhile, RocksDB must pay extra costs for compactions.

However, when facing workloads that regularly update existing keys
(with Zipfian and Zipfian-Composite distributions),
KVell shows significantly degraded throughput and up to $8.5\times$
more data written to the SSD compared with FlexDB.
The reason is that KVell uses slab allocators to manage space in the SSD and
must perform block-sized in-place updates, which leads to high WA
when the average KV size is smaller than the block size.
FlexDB shows higher throughput than RocksDB by 2.2--3.3$\times$ across the experiments,
The advantage mainly comes from FlexDB's capability of
directly committing updates to the FlexSpace at low cost.
In contrast, RocksDB requires repeated compactions to sort-merge KV pairs
across the multi-level structure, which leads to high WA and computation cost.
As shown in Figure~\ref{fig:flexdb-micro-wa},
RocksDB writes 2.1--2.9$\times$ more data to the SSD than FlexDB.

\paragraph{Read (GET and SCAN)}
We measure the point and range query
throughput of the three systems.
For each dataset, we populate the store with 4 threads,
followed by 4\,GB of random updates using the Zipfian distribution
to emulate a randomized data layout in real-world KV stores.

As shown in Figure~\ref{fig:flexdb-micro-read}, RocksDB shows low \texttt{GET} throughput because each operation requires a number of key comparisons to identify candidate tables at each level.
For each candidate table, it needs to examine the bloom filter and then search the index if the filter returns true.
KVell and FlexDB achieve higher throughput by maintaining
a single-level in-memory index for fast lookups.
The advantage of FlexDB is particularly high because it uses
a much smaller sparse index
and can quickly search in an interval with few key comparisons (see \S\ref{sec:db-cache}).
Additionally, KVell stores the block address of each KV pair in the full index.
A lookup in KVell needs to retrieve the cached block with an extra lookup in the page cache,
which adds a constant overhead.

As shown in Figure~\ref{fig:flexdb-micro-scan},
the advantage of FlexDB remains significant in range queries
because of its low cost on accessing KV data in the interval cache.
In comparison, range queries in RocksDB require expensive sort-merging of KV data
from multiple overlapping tables.
To avoid synchronization overhead,
KVell partitions the store with hash-based sharding,
where each shard is exclusively managed by a worker thread.
A range query in KVell must access every shard and
sort-merge all the KV pairs at the client side to generate the search results.
As a result, the scans are bottlenecked by
excessive data copying and sort-merging.

\paragraph{Scalability}
To measure the scalability of FlexDB, we rerun the write and read experiments
with 1 to 8 client threads using the UDB dataset and the Zipfian access pattern.
The scan experiments use a scan length of 50 keys.
The results are shown in Figure~\ref{fig:flexdb-scale}.
FlexDB and RocksDB both scale well in the read (and also write for RocksDB) experiments
because the workloads are mainly CPU-bound.
However, FlexDB's write throughput stops increasing with more than 5 threads.
In this scenario, the committer thread in FlexDB has been fully loaded
and becomes the bottleneck.
KVell shows constant throughput because it has a fixed number of worker threads,
each exclusively processing requests for a shard.
We reconfigure KVell with different numbers of shards, and
the \texttt{GET} performance reaches its peak at 1.96\,Mops/sec with
8 worker threads and 2 client threads (on the 10-core machine).
The \texttt{PUT} and \texttt{SCAN} throughput do not improve
since the I/O bandwidth is already saturated with four workers.

\paragraph{Latency}

\setlength{\tabcolsep}{2.1pt}
\begin{table}[h]
\caption{Latency and Throughput with UDB+Zipfian}
\begin{tabular}{c||c|c|c|c||c|c|c|c}
\hline
Op. & \multicolumn{4}{c||}{PUT} & \multicolumn{4}{c}{GET} \\ \hline
Sys. & Rocks & KVell & KVell$_{1}$ & Flex & Rocks & KVell & KVell$_{1}$ & Flex\\ \hline
Avg. ($\mu$s)  & 13.8 &    1669 &   153 & 3.9
               &  9.0 &     453 &  72.6 & 3.7 \\ \hline
95\,p ($\mu$s) &   17 &    2904 &   271 &   9
               &   21 &    953  &   143 &   9 \\ \hline
99\,p ($\mu$s) &   19 &    3386 &   306 &  17
               &   43 &    1360 &   173 &  33 \\ \hline
Mops/sec  & 0.30 & 0.53 & 0.09 & 0.95
          & 0.52 & 1.13 & 0.15 & 1.65 \\ \hline
\end{tabular}
\label{tab:latency}
\end{table}

We discuss the latency metrics with the UDB dataset under Zipfian workloads
(shown in Table~\ref{tab:latency}).
Compared with RocksDB, FlexDB is able to quickly commit KV updates to the FlexSpace instead of
merging data in a multi-level structure.
In the meantime, a lookup in FlexDB does not need to access multiple tables and sort data on the fly.
Therefore, FlexDB shows the lowest latency metrics in both \texttt{PUT} and \texttt{SET} operations.
KVell relies on asynchronous I/O to gain high throughput with a deep request queue (up to 64 queued requests).
The queuing causes much longer response times than in FlexDB and RocksDB.
That said, a smaller queue depth can improve responsiveness and reduce the latency readings of KVell.
Accordingly, we measure KVell's latency metrics with its queue depth set
to 1 and show the results in the columns named ``KVell$_1$'' in
Table~\ref{tab:latency}.
KVell's latency metrics improve by about an order of magnitude
by reducing the queue depth from 64 to 1, but the absolute numbers are still worse than FlexDB and RocksDB.
Furthermore, the improvement comes at a cost of mediocre throughput because of the lack of I/O parallelism, as shown in the last row in Table~\ref{tab:latency}.

\paragraph{GC overhead}

We evaluate the impact of the FlexSpace GC activities on FlexDB
using an update-intensive experiment.
Each run of the experiment performs in total 800 million
KV updates to a store containing the UDB dataset.
The total update size is approximately twice the store size,
which generates a fully aged storage layout during the experiment.
We first run the experiment with the FlexSpace's data file size capped at 128\,GB,
which represents the scenario of a modest space utilization ratio ($50\%$) and low GC overhead.
For comparison, we run the same experiments with the data file size capped at 75\,GB.
The smaller size leads to high GC activities in the FlexSpace with a higher space utilization ratio (85\%).
The results are shown in Figure~\ref{fig:flexdb-gc}.
The intensive GC shows a negligible impact on both throughput and I/O with Zipfian workloads.
In this scenario, the GC process can easily find near-empty segments
because the frequently updated keys are often co-located in the data file.
Comparatively, the Zipfian-Composite distribution has a much weaker spatial locality,
which leads to more rewrites in the GC process.

\paragraph{YCSB Benchmark}

\begin{figure}[h]
    \centering
    \begin{subfigure}{0.26\textwidth}
        \centering
        \includegraphics[height=2.9cm]{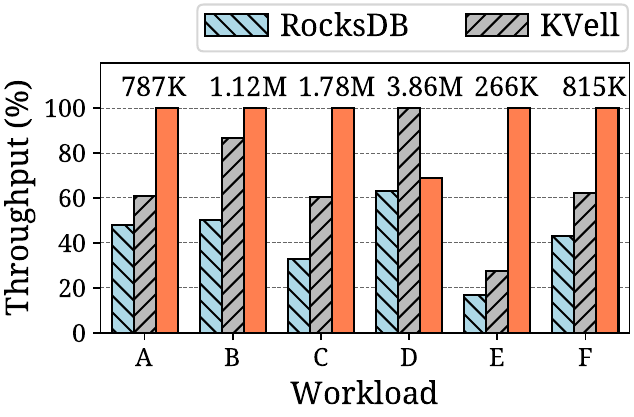}
        \caption{64\,GB store}
        \label{fig:flexdb-ycsb}
    \end{subfigure}
    \begin{subfigure}{0.21\textwidth}
        \centering
        \includegraphics[height=2.9cm]{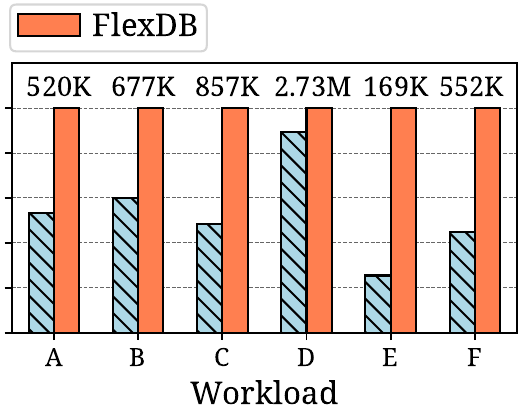}
        \caption{500\,GB store}
        \label{fig:flexdb-ycsb-oc}
    \end{subfigure}
    \caption{YCSB benchmark with the UDB KV data sizes.
    Results of each workload are normalized to the highest.}
\end{figure}

YCSB~\cite{cooper2010ycsb} is a popular benchmark that evaluates KV store
performance using realistic workload patterns.
We use the UDB store populated by the corresponding four-thread load experiment,
and run the YCSB workloads from A to F.
The details of the YCSB workloads are shown in Table~\ref{tab:ycsb}.
A scan in workload~E performs a seek and retrieves 50 KV pairs.
Figure~\ref{fig:flexdb-ycsb} shows the benchmark results.

\setlength{\tabcolsep}{4pt}
\begin{table}[h]
\centering
\caption{YCSB workloads}
\begin{tabular}{l|c|c|c|c|c|c}
\hline
Workload & A & B & C & D & E & F \\ \hline
Distribution &
\multicolumn{3}{c|}{Zipfian} & Latest & \multicolumn{2}{c}{Zipfian} \\ \hline
\multirow{2}{*}{Operations} &
50\% U & 5\% U & \multirow{2}{*}{100\% R} & 5\% I & 5\% I & 50\% R \\
& 50\% R & 95\% R & & 95\% R & 95\% S & 50\% M \\ \hline
\multicolumn{7}{l}{$^*$ \footnotesize I: Insert; U: Update; R: Read; S: Scan; M: Read-Modify-Write.} \\
\end{tabular}
\label{tab:ycsb}
\end{table}

In read-dominated workloads including B, C, and E, FlexDB outperforms RocksDB and KVell
by $2.0$--$5.9\times$ and $1.2$--$3.6\times$, respectively.
This is especially the case in workload~E because of FlexDB's advantage in range queries.
Workload D performs sequential write while reading
very recent updates with an ideal access locality.
KVell achieves the highest throughput because it can evenly distribute requests
across the hash-based shards without lock contention.

In write-dominated workloads, including A and F, FlexDB outperforms RocksDB and KVell
by $2.1$--$2.3\times$ and $1.6\times$, respectively.
The performance advantage is not as high as that in the read-dominated workloads.
In the FlexDB implementation,
when the committer thread is merging updates into the FlexSpace,
readers that reach the sparse index can be temporarily blocked (see \S\ref{sec:db-co}).
In workload A, the P99 latency is 30\,$\mu$s with a maximum reader blocking time of 3.4\,ms.
The blocking time can be improved by partitioning the store~\cite{gilad2020evendb},
which is beyond the scope of this paper.

We also run the YCSB benchmark in an out-of-core scenario
by increasing the UDB dataset size to about 500\,GB (3.2 billion keys).
In this setup, KVell's full index does not fit in the available RAM.
When running with swap space enabled,
KVell shows severe performance degradation by more than an order of magnitude
compared to the in-core experiments, except for workload D that has optimal locality.
Similar slowdowns are also observed in KVell's evaluation on the impact of different memory sizes~\cite{lepers2019kvell}.
Therefore, we do not turn on swap and exclude KVell from this experiment.

Figure~\ref{fig:flexdb-ycsb-oc} shows the out-of-core benchmark results.
In this scenario,
both FlexDB and RocksDB show reduced throughput in all the YCSB workloads
due to the increased I/O cost.
The advantage of FlexDB over RocksDB is reduced in the most I/O-intensive workloads
(C and E).
This is because the increased I/O time overshadowed FlexDB's search efficiency on the sparse index.
That said, FlexDB still achieves 1.1--3.9$\times$ speedups over RocksDB.

\paragraph{Recovery}

We evaluate FlexDB's recovery speed (described in \S\ref{sec:db-recovery})
with a clean page cache and four concurrent recovery threads.
For a store containing the 64\,GB UDB dataset,
the recovery process takes 7.8s using a small rebuilding interval size of 16\,KB.
Increasing the recovery interval size to 64\,KB
reduces the recovery time to only 1.9s.
In practice, users can make trade-offs between reduced service downtime and
better first-time access latency by adjusting the recovery interval size.
Besides, the first-time access latency can be further reduced
by promptly warming up the intervals in the background using spare bandwidth.
RocksDB also achieves fast recovery by only scanning the WAL and lazily loading table files
on demand.
In comparison, KVell uses 64 seconds to rebuild a full
index in the memory with four worker threads,
and a complete scan of all the keys is inevitable in this process because of the unordered persistent storage layout of KVell.
\section{Related Work}
\label{sec:related-work}
\paragraph{Data-management Systems}

Studies on improving I/O efficiency in data-management
systems are abundant~\cite{zhang2015data, davoudian2018survey}.
B-tree-based KV stores~\cite{lmdb, olson1999berkeley, mongodb}
support efficient searching with minimum read I/O
but have suboptimal performance under random writes because of
the in-place updates~\cite{li2010fdtree}.
LSM-Tree~\cite{oneil1996lsmtree} uses out-of-place writes
and delayed sorting to improve write performance,
and it has been widely adopted 
in write-optimized KV stores~\cite{rocksdb, leveldb}.
However, the improved write efficiency comes at a cost of
slow read operations since a search may
query multiple tables at different locations~\cite{luo2020lsm}.
To compensate reads, LSM-tree based KV stores need to rewrite table files
periodically using a compaction process,
which in turn offsets the benefit of
out-of-place write~\cite{raju2017pebblesdb,ren2017slimdb,wu2015lsm,dayan2019wacky,dayan2018dostoevsky,im2020pink,bortnikov2018accordion,huang2019xengine}.
KVell and HiKV index all the keys
in a volatile ordered index for fast access and leaves KV data unsorted on the persistent storage~\cite{lepers2019kvell,xia2017hikv,chandramouli2018faster}.
However, maintaining a volatile full index leads to high memory footprints
and lengthy recovery processes.
SplinterDB employs B$^\varepsilon$-tree
for fast write by logging unsorted KV pairs in tree nodes~\cite{conway2020splinterdb}.
However, the unordered node layout leads to slow reads,
especially for range queries.
Hashing-based KV stores gain point query efficiency but have to
give up support to range queries~\cite{wu2015lsm,kourtis2019udepot}.
Recent studies also employ byte-addressable NVM for fast access and
persistence~\cite{kaiyrakhmet2019slmdb,yao2020matrixkv,kannan2018novelsm,chen2021spandb,benson2021viper}.
These solutions require non-trivial implementation,
including space allocation, GC, and maintaining crash consistency,
which overlaps the core duties of file systems.
FlexDB delegates the challenging data organizing tasks
to the mechanisms behind the persistent address space,
which effectively reduces application complexity.
Managing persistently sorted KV data with efficient in-place updates
achieves fast read and write at low cost.

\paragraph{Address Space Management}
Modern in-kernel file systems, such as Ext4, XFS, Btrfs, and F2FS,
use B$^+$-Tree and its variants or multi-level mapping tables
to index file extents~\cite{htree, xfs, rodeh2013btrfs,lee2015f2fs}.
These file systems provide comprehensive support for general file management tasks
but exhibit suboptimal performance in metadata-intensive workloads,
such as massive file creation, crowded small writes,
as well as \textit{insert-range} and \textit{collapse-range} that require data shifting.
Recent studies employ write-optimized data structures in file systems to improve
metadata management performance.
Specifically, BetrFS~\cite{jannen2015betrfs,yuan2016optimizing,zhan2018fullpath},
TokuFS~\cite{esmet2012tokufs}, WAFL~\cite{macko2010wafl},
TableFS~\cite{ren2013tablefs}, and KVFS~\cite{shetty2013kvfs}
use write-optimized indexes, including B$^\varepsilon$-Tree~\cite{bender2015betree}
and LSM-Tree~\cite{oneil1996lsmtree}, to manage file system metadata.
Their designs exploit the advantages of these indexes and
successfully improved many existing file system metadata and file I/O operations.
However, these systems still employ the traditional file abstraction
and do not support easily moving data in the file address space.
Therefore, rearranging file data in these systems still relies on rewriting existing data.
In-memory systems such as rewired memory~\cite{schuhknecht2016ruma, leo2019pmar}
utilize virtual memory mappings (i.e., page tables) to dynamically relocate page-aligned in-memory data blocks to sort data without copying.
These mechanisms suffer from the same data shifting problems as in file extent indexes.
The design of FlexSpace removes a fundamental limitation
in persistent address spaces.
By leveraging the efficient shift operations for logically reorganizing data,
applications built on FlexSpace can easily avoid data rewriting in the first place.
\section{Conclusion}
\label{sec:conclusion}

This paper presents a novel storage engine that provides a \textit{flexible address space},
which enables lightweight and efficient in-place updates.
It allows applications to perform efficient data management on a linear data
layout with a simplified implementation.
FlexDB, a KV store built on FlexSpace with a simple structure,
achieves speedups of up to $16\times$ for read and 3.3$\times$ for write,
compared with highly optimized KV stores.




\printbibliography


\end{document}
\endinput